\documentstyle[aps]{revtex}

\begin{document}
\draft
\preprint{}

\title{
Flux Pinning and Phase Transitions in Model High-Temperature\\
Superconductors with Columnar Defects}

\author{K. H. Lee and D. Stroud}
\address{Department of Physics, Ohio State University, Columbus,
Ohio 43210, USA}

\author{S. M. Girvin}
\address{Department of Physics, Indiana University, Bloomington,
Indiana 47405, USA}

\date{\today}
\maketitle

\begin{abstract}
We calculate the degree of flux pinning by
defects in model high-temperature superconductors (HTSC's).  The HTSC is
modeled as a three-dimensional network of resistively-shunted Josephson
junctions in an external magnetic field, corresponding to a HTSC in the
extreme Type-II limit.  Disorder is introduced either by
randomizing the coupling between grains (Model A disorder) or by removing
grains (Model B disorder). Three types of defects are
considered: point disorder,
random line disorder, and periodic line disorder;
but the emphasis is on random line disorder.  Static and dynamic
properties of the models are determined by Monte Carlo simulations and by
solution of the analogous coupled overdamped Josephson equations in the
presence of thermal noise.   Random line defects
considerably raise the superconducting transition temperature T$_c(B)$, and
increase the apparent critical current density J$_c(B,T)$,
in comparison to the defect-free crystal.  They are more effective in
these respects than a comparable volume density of point defects, in
agreement with the experiments of Civale {\it et al}.  Periodic line
defects commensurate with the flux lattice are found to raise T$_c(B)$ even
more than do random line defects.  Random line defects are most effective
when their density approximately equals the flux density.  Near
T$_c(B)$, our static and dynamic results appear consistent with
the anisotropic Bose glass scaling hypotheses of Nelson and Vinokur, but
with possibly different critical indices:
transverse correlation length exponent $\nu_{\perp} \approx 1.2$, anisotropy
exponent z $\approx 1.4 \pm 0.2$
(where z is defined by $\xi_{\|} \propto \xi_{\perp}^z$, with $\xi_{\|}$
and $\xi_{\perp}$ are the correlation lengths parallel and perpendicular to
the flux), and dynamical critical exponent z$^{\prime} \approx 2.0$.
\end{abstract}


\narrowtext
\twocolumn

\section{Introduction.}

A major problem restricting the practical use of high-temperature
superconductors (HTSC's)
is the difficulty of producing a large critical current,
especially in a magnetic field.$^{1}$
Much of this difficulty is thought to result from dissipation due to flux
motion - a dissipation generally known at low or high dissipation rates
as ``flux creep''$^{2-6}$ or ``flux flow''$^7$ respectively.  When a
current density $\vec{J}$ is introduced into the HTSC, it produces a
$\vec{J} \times \vec{B}$ force (known as a Magnus force) on the flux lines.
This force tends to set the lines in motion, producing resistive
dissipation, unless appropriate defects, known as pinning centers, can
prevent this motion, or at least raise the current density at which it
begins.

Recently, Civale {\it et al},$^8$ in an elegant set of experiments, have shown
that columnar defects, introduced
parallel to the flux lines by heavy ion irradiation, can greatly
increase the critical current at which flux motion dissipation begins,
relative to the point defects which are more commonly introduced as pinning
centers, e. g. by proton irradiation$^{9-11}$  The same columnar defects were
also found to increase the
temperature of the so-called ``irreversibility line''$^{12}$ in the magnetic
field-temperature (H-T) plane, below which flux motion essentially ceases
in the limit of a weak applied current.  The columnar pins were produced by
irradiating the HTSC with a beam of heavy ions parallel to the c-axis.  It
is not surprising that columnar defects should be effective pins:
they provide a long pinning center which should provide a much stronger
pinning potential for a long flux line than will an equal concentration of
point defects.  However, a realistic calculation which demonstrates this
effect has been lacking.

In this paper, we present some simple model calculations which demonstrate
both of the effects observed by Civale {\it et al},$^8$ and also suggest some
alternative methods for further increasing both the critical current and
irreversibility temperature of HTSC's.  Our approach is to describe the
HTSC as a three-dimensional collection of resistively-shunted Josephson
junctions (RSJ's), in which temperature is simulated by a Langevin noise
source of the appropriate strength in each junction.$^{13}$  Such a model is
obviously far from a realistic HTSC.  However, the model does
contain some of the essential physics needed to describe transport in HTSC's:
it embodies coupled fluctuating phases within the context of a reasonable
dynamics, and it allows for the introduction of an applied magnetic field
in a simple way.  In this view, the Josephson-coupled ``grains'' should
probably be considered as representing
small patches of phase-coherent superconductor, of
dimensions comparable to the coherence length.$^{14}$  Thus, the model is not
restricted to literally granular materials, but could apply to single
crystals with more microscopic disorder, in the extreme type-II limit
(penetration depth $\lambda$ much larger than coherence length $\xi$)..
We have shown elsewhere that a similar model
describes the difference between transverse and longitudinal
magnetoresistance of a HTSC, in qualitative agreement with experiment.$^{15}$

We turn now to the body of the paper.  Section II describes the model for
both the thermodynamic and transport properties of the HTSC with defects.
Section III describes our numerical results for these properties.  A brief
discussion follows in Section IV.  Three Appendices summarize the static
and dynamic scaling hypotheses used to analyze our numerical results.

\section{Model.}

\subsection{Thermodynamics.}

We consider a simple cubic three-dimensional network of N superconducting
``grains''
weakly coupled together by Josephson junctions, and driven by an externally
applied current.  The i$^{th}$ ``grain'' is described by a superconducting
order parameter $|\psi_i|\exp(i\theta_i)$.  We neglect fluctuations in the
amplitude $|\psi_i|$ and allow the phase $\theta_i$ to fluctuate.  With
these assumptions, the {\it thermodynamic} properties of the
network are given by the Hamiltonian
\begin{equation}
H = -\sum_{<ij>}E_{J;ij}\cos(\theta_i-\theta_j-A_{ij})
\end{equation}
where
\begin{equation}
E_{J;ij} \equiv \frac{\hbar}{2e}I_{c;ij}
\end{equation}
is the coupling energy between grains i and j, I$_{c;ij}$ is the
corresponding critical current,
\begin{equation}
A_{ij} = \frac{2\pi}{\Phi_0}\int_i^j\vec{A}\cdot \vec{dl},
\end{equation}
$\Phi_0=hc/2e$, and $\vec{A}$ is the vector potential, taken to be that
of the externally applied magnetic field (this is equivalent to assuming
a Josephson penetration depth large compared to the intergranular
separation).  The sum runs over distinct nearest neighbor pairs.

Given H, equilibrium properties are obtained via an average with respect to
to a canonical ensemble.  Thus, for example, the average of some operator
$O(\theta_1,...\theta_N)$ is obtained from
\begin{equation}
<O> = \int O(\theta_1,...,\theta_N)e^{-H(\theta_1,...,\theta_N)/k_BT}
\Pi_{i=1}^Nd\theta_i/Z
\end{equation}
where
\begin{equation}
Z = \int \Pi_i d\theta_i \exp(-H/k_BT).
\end{equation}
is the canonical partition function.

In the calculations to be described below, we have generally dealt with
{\it disordered} samples.  In that case, we calculate
averages both over a canonical ensemble (denoted $<...>$) and over
different realizations of the disorder (denoted [...]).  We have considered
primarily the specific heat per grain $C_V$ and the so-called helicity
modulus tensor with components $\gamma_{ij}$.  C$_V$ is generally computed
from the fluctuation expression
\begin{equation}
C_V = [<H^2>-<H>^2]/(Nk_BT^2).
\end{equation}  The helicity modulus (or equivalently, the superfluid
density) is the free energy cost of imposing a twist in the phase at the
boundaries of the sample.  Its principal elements are essentially spin-wave
stiffness constants.  Rather than imposing a twisted boundary condition and
calculating the resulting increase in free energy, it is more convenient to
use periodic boundary conditions and calculate
$\gamma_{ij}$ as
\begin{equation}
\gamma_{ij} = \left( \frac{\partial^2F}
{\partial A_i^{\prime}\partial A_j^{\prime}} \right)_{\vec{A}^{\prime}=0}.
\end{equation}
Here $\vec{A}^{\prime}$ represents an added uniform vector potential (in
addition to that which produces the applied magnetic field) which is
included in the Hamiltonian in order to produce a twist.  The various
second derivatives in eq. (7) are readily computed explicitly for an
ordered or a disordered sample, with the result for, e. g., $\gamma_{xx}$:
\begin{eqnarray}
N\gamma_{xx}=\left[\left<\sum_{<ij>}
E_{J;ij}x_{ij}^2\cos(\theta_i-\theta_j-A_{ij})\right>
\right]- \nonumber \\ -\frac{1}{k_BT}\left[\left<\left(
\sum_{<ij>}E_{J;ij}x_{ij}\sin(\theta_i-
\theta_j-A_{ij})\right)^2\right>\right]+ \nonumber \\
+\frac{1}{k_BT}\left[\left<\sum_{<ij>}
E_{J;ij}x_{ij}\sin(\theta_i-\theta_j-A_{ij})\right>^2\right].
\end{eqnarray}
where x$_{ij}=x_j-x_i$ is the x coordinate of the distance between nearest
neighbor grains i
and j.  Similar expressions hold for the other components of
$\gamma$.$^{16,17}$

\subsection{Dynamics.}

There are many dynamical models whose
{\it equilibrium thermodynamic properties} are represented by the
model just described.  We choose a dynamical model corresponding to
a simple cubic array of
overdamped resistively-shunted Josephson junctions, driven by an applied
current.  The network is then characterized by the set of coupled equations
\begin{eqnarray}
I_{ij} = I_c\sin(\theta_i-\theta_j-A_{ij}) + \frac{V_{ij}}{R_{ij}} +I_{L;ij} \\
V_{ij} \equiv V_i-V_j = \frac{\hbar}{2e}\frac{d}{dt}(\theta_i-\theta_j) \\
\sum_j I_{ij} = I_{i;ext}\\
A_{ij} = \frac{2\pi}{\Phi_0}\int_{\vec{x}_i}^{\vec{x}_j}\vec{A}\cdot\vec{dl}.
\end{eqnarray}
Here I$_{ij}$ is the current from grain i to grain j, which is written as
the sum of a Josephson current and an Ohmic current through the shunt
resistance R$_{ij}$; V$_{ij}$ is the voltage difference between
grains i and j; and I$_{i;ext}$ is the external current fed into grain i.
In the calculations described below, the current is always fed into one
face of the array (an equal amount I into each grain)
and extracted from the opposite face, with periodic
transverse boundary conditions.  Eq. (11) is
Kirchhoff's law describing current conservation at grain i. I$_{L;ij}$ is a
Langevin noise current$^{18}$ between grains i and j, introduced to simulate
the
effects of temperature, which satisfies the relation
\begin{eqnarray}
<I_{L;ij}(t)> = 0 \\
<I_{L;ij}(t)I_{L;kl}(t')> = \frac{2k_BT}{R_{ij}}\delta_{ij;kl}\delta(t-t'),
\end{eqnarray}
where T is the absolute temperature and the brackets denote an ensemble
average.  We solve these equations by Euler iteration, as described
previously.$^{13,15}$

\subsection{Geometrical Models for Line and Point Disorder.}

We have considered two types of models to describe disorder, which we
denote Model A and Model B.  In Model A, the bond energy
E$_{J;ij}$ between grains i and j is assumed to vary randomly between 0 and
twice its average value.  In Model B, we introduce disorder simply by
removing a certain fraction of the grains, as well as their associated
Josephson junctions (but not shunt resistances).

For either Model A and Model B, we can assume either ``line disorder''
or ``point disorder.''  In Model A, point disorder can be introduced by
assuming
that the bond strengths of different bonds are completely uncorrelated.
``Line disorder'' consists of assuming that the
bond strengths are uncorrelated in the xy plane, but are perfectly
correlated in the z direction - that is, the strength of a given bond,
whatever its orientation, depends the x and y coordinates describing its
location, but not on the
z coordinate.

To introduce point disorder in Model B, we remove the grains at random.  For
line disorder, we remove lines of grains parallel to the z axis.  The
removal of these grains effectively converts the neighborhood of the grain
from superconducting to normal.  In Model B, we have also considered
``periodic line disorder,'' in which line defects are arranged periodically
in the xy plane, as described further below.

\subsection{Magnetic Field.}

We consider magnetic fields applied in
the z-direction, i.e., parallel to the line defects.  In previous
calculations of this sort, it has been standard to use the Landau gauge,
$\vec{A}=Bx\hat{y}$.   This gauge severely restricts the possible magnetic
fields that can be considered if one also requires periodic boundary
conditions in all three directions (as in the Monte Carlo simulations) or
in the transverse directions (as with the dynamic simulations).
We therefore use a different gauge previously used by Arovas and Haldane in
other contexts$^{19}$.

To define this gauge, we consider an L $\times$ L square array of lattice
constant $a$, with the origin taken as the lower left hand corner of the
array.  Then we take
A$_{ij} = 2\pi fn$ for bonds pointing in the y direction and
located at x = na;
A$_{ij}$ = 0 for all bonds in the x direction except for those in the
extreme right-hand column of plaquettes, and A$_{ij}$ = -2$\pi fLm$ for
horizontal
bonds in that extreme right-hand column, at y = ma, x=La.  (In this expression
f = $\Phi/\Phi_0$, where $\Phi$ is the flux per plaquette.)  With this
choice of gauge, it is readily verified that the factors
A$_{ij}$ sum to 2$\pi f$ (modulo 2$\pi$)
around each plaquette, as required.  The
requirement of periodicity in the two transverse directions will now be
satisfied as long as f is a multiple of 1/L$^2$.
This is much weaker than the condition imposed by periodicity when the
Landau gauge is used, which is that f be a multiple of 1/L.  Hence, a much
finer grid of magnetic fields can be considered than in most previous
calculations.

\section{Numerical Results}

\subsection{Model A.}

We begin by presenting our numerical results for Model A.  Fig. 1 shows the
specific heat C$_V$ per grain and the helicity moduli $\gamma_{\|}$ and
$\gamma_{\perp}$ parallel and perpendicular to the magnetic field, for an
{\it ordered} array of size L $\times$ L $\times$ L grains with periodic
boundary conditions, isotropic coupling, magnetic field f = 1/4, and
several different sizes (L = 8, 12, and 16).
$C_V$ shows clear signs of diverging in the vicinity of k$_B$T = 1.1E$_J$
(the peak in C$_V$ is growing with increasing lattice size, suggesting a
continuous phase transition).  We interpret
this temperature as a melting transition with a discrete symmetry associated
with the periodic lattice.  Similar numerical results have been obtained for
this model by Shih {\it et al}$^{17,20}$.

We may get a clearer quantitative picture of this phase transition by
applying a static scaling analysis,$^{21-23, 16}$
as described in detail in Appendix A.
In Fig. 2, we plot
L$\gamma_{\|}$ and L$\gamma_{\perp}$ for several values of L as a function
of temperature.  As is clear from the Figure,
each attains a universal value near the same
temperature k$_B$T$_c$ = 1.1E$_J$, suggesting that this is indeed the
critical temperature for this model.
When combined with
the scaling analysis of Appendix A, these results suggest that
this phase transition is characterized by a {\it single} correlation
length $\xi$, i.e. z=1, where z is the anisotropy exponent,
defined by the relation
$\xi_{\|} \propto \xi_{\perp}^z$, where $\xi_{\|}$ and $\xi_{\perp}$ are the
correlation lengths parallel and perpendicular to the field.

Fig. 3 shows the quantities of Fig. 1, but with
{\it line disorder},$^{24}$
Once again, we consider isotropic coupling strengths and
use f = 1/4, and several different (but cubic) box sizes.
The calculations shown are the
result of averages over many realizations of the disorder, as indicated in
the legends of the Figure.  In
contrast to Fig. 1, there is very little size dependence of C$_V$,
suggesting that C$_V$ either does not diverge or at most diverges very
weakly.  Note also that the helicity modulus $\gamma_{\|}$ goes
to zero at a substantially higher temperature than in the ordered case,
indicating that the superconducting transition temperature is
{\it increased} by the line disorder.

Another point is that, like the ordered case,
$\gamma_{\|}$ also seems to vanish {\it continuously} with
temperature.  If we assume a power law of the form
$\gamma_{\|} \propto |T-T_c|^{g_{\|}}$, then Fig. 3 suggests
g$_{\|} \approx 0.5$ and $k_BT_c \approx 1.7<E_J>$  The
the numerical uncertainties in $\gamma_{\perp}$ are much larger, so an
accurate estimate of the analogous quantity $g_{\perp}$ is difficult.
Indeed, at any given temperature,
the Monte Carlo convergence of $\gamma_{\perp}$ is much slower than for
$\gamma_{\|}$.  This is presumably
because our model is both frustrated and disordered
in the xy plane, but is neither frustrated nor disordered in the z
direction.  Hence, the system rapidly responds to any twist in that
direction, but much more slowly in the xy plane.  Also,
$\gamma_{\perp}$ converges more slowly in the disordered case than in the
frustrated but ordered model of Fig. 1, suggesting that the slow
convergence is caused by the
huge number of metastable states of nearly equal energy which are expected
in the disordered system.

{}From the static scaling analysis of Appendix A, we can roughly estimate
the anisotropy exponent z defined there.$^{21}$
{}From eq. (23) of Appendix A, $g_{\|} = \nu_{\perp}(2-z)$, where
$\nu_{\perp}$ is related to the transverse correlation length $\xi_{\perp}$
by $\xi_{\perp} \propto |(T-T_c)/T_c|^{-\nu_{\perp}}$.  Hence,
our numerical results suggest
\begin{equation}
\nu_{\perp}(2-z) \approx 0.5.
\end{equation}
If $\nu_{\perp}$ is finite, this equality suggests that z $<$ 2 for this model.
Secondly, since C$_V$ is apparently nondivergent,
eq. (21) shows that
\begin{equation}
(2+z)\nu_{\perp} \geq 2.
\end{equation}
Adding these results, we get $4\nu_{\perp} \geq 2.5$ or
$\nu_{\perp} \geq 0.63$.
Hence, $(2-z)=0.5/\nu_{\perp} \leq 0.8$ or
$z \geq 1.2$.  Combining all these arguments, we estimate $1.2 \leq z \leq 2$.

In order to go further, we must estimate $g_{\perp}$.
According to eq. (24) of Appendix A, $g_{\perp} = z\nu_{\perp}$.  From
Fig. 3, we estimate z$\nu_{\perp} \approx 1-2$.  Combining this with eq. (15),
and assuming z$\nu_{\perp} =1.5$, we get z $\approx$ 1.5,
$\nu_{\perp} \approx 1$, which satisfy the inequalities
described above.  They also agree with
Chayes {\it et al}$^{25}$, who have proposed for a wide class of
continuous-spin models with disorder, that
$\nu_{\perp} \geq 1$ rigorously.  Our estimate is, of course,
subject to large numerical uncertainties, especially in g$_{\perp}$.

We turn now to the dynamical properties of Model A, concentrating on
isotropic coupling with and without line disorder.
Fig. 4 shows the IV characteristics of a 6 $\times$ 6 $\times$ 9 array
at magnetic field f = 1/4, with no defects and with current
density $\vec{J} \perp \vec{B}$, plotted at several different temperatures.
Figs. 5(a) and 5(b) show the IV characteristics for the analogous model with
line disorder, a 6 $\times$ 6 $\times$ 6 unit cell, and two current
orientations: $\vec{J} \perp \vec{B}$ and $\vec{J} \| \vec{B}$.
Like the Monte Carlo results, the IV characteristics
also suggest that T$_c$ is increased by line defects.  To make this clearer,
we have plotted in Fig. 6 the {\it resistivity} $\rho \equiv <V>/(LRI)$ at a
current level I=0.05I$_c$: the introduction of line defects
{\it reduces} $\rho$ at every temperature,
relative to the no-defect case.

The {\it shape} of the IV characteristics is also changed by the
introduction of line defects.  With no line defects and
$\vec{J} \perp \vec{B}$ (Fig. 4), there is a
fairly clear critical current onset for temperatures $k_BT < 1.0E_J/k_B$.
When line defects are present, the IV characteristics suggest no clear
critical current for $\vec{J} \perp \vec{B}$
(the analytic form of these IV characteristics is discussed further below).
By contrast, for $\vec{J} \| \vec{B}$,
a critical current seems to develop for  $k_BT < 1.3-1.4E_J$.

We have attempted to scale the IV characteristics of Fig. 5
according to the formalism$^{21}$ outlined in Appendix B.
For $\vec{J}\perp\vec{B}$, we plot
E$_{\perp}t^{-\nu_{\perp}(1+z')}$ against $J_{\perp}t^{-\nu_{\perp}(1+z)}$
(where t = $|T-T_c|/T_c$) for various estimates of T$_c$, z,
z$^{\prime}$,and $\nu_{\perp}$.  For $\vec{J} \| \vec{B}$, we plot
E$_{\|}t^{-\nu_{\perp}(z+z')}$ against $J_{\|}t^{-2\nu_{\perp}}$.
Our best results are shown in Figs. 7(a) and 7(b).  For a given choice of
T$_c$, the best fits seem to correspond to somewhat different values of the
parameters in the parallel and perpendicular directions -
not surprising in view of the numerical uncertainties, small
sample sizes, and limited current ranges.  For $\vec{J}\perp\vec{B}$,
assuming k$_B$T$_c = 1.7E_J$, we get $\nu_{\perp}(1+z)\approx3.8$,
$\nu_{\perp}(1+z^{\prime})\approx 4.5$. For $\vec{J}\|\vec{B}$, on the
other hand, we get 2$\nu_{\perp}\approx 2.4$,
$\nu_{\perp}(z+z^{\prime})\approx 3.4$.  Combining these results with the
previous Monte Carlo estimates, we estimate z = $1.4\pm 0.2$,
z$^{\prime} = 1.8\pm 0.3$, and $\nu_{\perp} = 1.3\pm 0.3.$
The error estimates in all three cases are simply subjective assessments of
our confidence in these numbers over the extremely limited current and size
ranges considered.   This estimate is also based on the assumption that the
{\it same} value of $z^{\prime}$ applies to both the parallel and
perpendicular case.  That assumption is apparently {\it not} valid in
the case of short-range interactions between the vortex lines, as has
recently been shown by Wallin and Girvin.$^{26}$

Both above and below the assumed T$_c$, the IV characteristics in both the
parallel and perpendicular case collapse adequately
(though not perfectly) onto universal scaling functions above and
below T$_c$.  For both current directions, but especially for
$\vec{J} \| \vec{B}$, there is a conspicuous
ohmic tail in the IV characteristics below T$_c$.  We believe this is a
finite-size effect, as further analyzed in
Appendix C.  Indeed, we have checked
that the tail becomes smaller and smaller as the size of the array is
increased.  When T $< T_c$, the perpendicular and parallel
scaling functions are quite dissimilar.  In both cases, we can obtain
acceptable fits with scaling functions of the form
E(x) $\propto$ exp$[-(A/x)^{\mu}]$, but the best fit for
$\mu_{\perp}$ is in the range 0.2-0.4 while $\mu_{\|} \approx 1$.  This is
consistent with the IV characteristics of Fig. 5, which show a much clearer
critical current developing in the parallel direction than in the
perpendicular direction.

\subsection{Model B Disorder.}

To study Model B disorder, we consider an
8 $\times$ 8 $\times$ 5 lattice with flux per plaquette f = 1/8, parallel
to the z (thin) direction.
 We have considered three types of defect configurations: (i) ``point
defects,'' introduced by removing at random 40 grains, and their associated
Josephson couplings (but not shunt resistances) from among the 320 grains of
the
lattice; (ii) ``random line defects,'' consisting of eight line defects,
each five grains long, parallel to the z direction but randomly distributed
in the xy plane; and (iii) ``periodic line defects,'' in which the line
defects are arranged with the periodicity of the ground state phase
configuration at f = 1/8.  For reference, we also consider (iv)
the no-defect configuration.

In the absence of an applied current, the phases of configuration (iv) will
settle into a z-independent ground state configuration.  This
can be found numerically, e. g., by starting the phases in a
random arrangement and iterating the Josephson equations at zero applied
current until a state of no voltage is obtained (care must be exercised to
avoid falling into a metastable minimum).   To calculate the IV
characteristics, we typically begin with this ground state,
gradually increasing the applied current for various defect
configurations.

Fig. 8 shows the resulting IV characteristics (for $\vec{J} \perp \vec{B}$)
at temperature T = 0.  Case (iv) has a critical
current $\approx$ 0.12I$_c$ per junction, comparable to
the calculated depinning critical current for a {\it single} vortex in a
large square array$^{27}$.  This suggests that the critical current is
not too much influenced, in this case, by vortex-vortex interactions.  The
critical current is increased slightly by point defects [case (i)],
somewhat more by random line defects [case (ii)],
and considerably more again by periodic line defects [case (iii)].
In case (ii), the {\it functional form} of the IV
characteristic is considerably modified by the defects (being concave up
rather than concave down).  We find that it is fairly well fitted by
$<V>/(LRI_c) = A$exp$[-C(I_c/I)^{\mu}]$ with $\mu \approx 0.3$,
A $\approx 3\times 10^4$, and $C \approx$ 10.

Fig. 9 shows the temperature-dependent resistivity $\rho(T) \equiv V/I$ at a
current of 0.1I$_c$ per junction, for cases (i) - (iv) and
$\vec{J} \perp \vec{B}$.  For reference, the superconducting transition
temperature T$_c(B=0)$
of the Hamiltonian (1) in zero magnetic field and zero current is known to
occur at k$_BT_c \approx 2.21E_J$.$^{28}$
Hence, Fig. 9 suggests that, whatever the defects,
T$_c$(f = 1/8) $<$ T$_c(f = 0)$ - that is, as expected, T$_c$ is reduced
in a magnetic field.  However, relative to the zero-defect case,
T$_c$(1/8) is increased slightly by point defects, more yet by random line
defects, and still more by periodically arranged line defects.
The increase in T$_c$(1/8) produced by random line defects corresponds to
the increase in the ``irreversibility temperature'' observed by Civale
{\it et al} when random line defects are introduced parallel to the
magnetic field.   Note that, in Fig. 9,  the density of line defects
``equals'' the density of point defects in the sense that an equal amount of
superconducting material is removed in each case.
The only difference among the various defect
curves in Fig. 9 is the degree to which
the disorder is correlated.  Hence, this plot
provides a very direct illustration of the influence of correlation in
raising the ``irreversibility temperature.''

Fig. 10 shows the T = 0 IV characteristics for a flux
density f = 1/8 and several densities f$_d$ of randomly distributed line
defects.  Fig. 11 shows a similar plot for $\rho(T)$ at a current level of
0.1 I$_c$.  These Figures suggest that (at least for f = 1/8)
both the T = 0 critical current and T$_c(B)$ are largest when
f $\approx$ f$_d$, a conclusions which may qualitatively agree with
experiment$^8$.  To confirm these conclusions, however, Monte Carlo
calculations should be carried out on the analogous Hamiltonian to
determine the dependence of T$_c$ on defect line density.

To summarize for Model B disorder, our results
show that columnar defects oriented parallel to the flux lines tend to
increase the critical current, and to push up the superconducting
transition temperature $T_c(B)$, relative to the same
number of random point defects at the same field, in apparent agreement
with experiment.
We predict also that a {\it periodic} arrangement of line defects
commensurate with the defect-free flux lattice
will be even more effective in increasing both the critical current and
T$_c(B)$.  Finally, we have preliminary evidence that the
pinning (and T$_c$-enhancing) effects of random line defects are optimized when
the defect density is comparable to the flux density.

\section{Discussion.}

Nelson and Vinokur$^{21}$
have recently proposed a theory of the superconducting
transition in materials with correlated disorder.  When the density of line
defects is greater than the density of flux lines, their theory leads to a
phase transition from a high-temperature flux liquid into low-temperature
``Bose glass'' phase.  In their theory, the three-dimensional superconductor
maps onto a two-dimensional system in which the flux lines play the lines
of interacting Bose particles while the line defects become static point
defects.  When the density of line defects equals the density of flux
lines, Nelson and Vinokur
predict instead a transition into a ``Mott insulator''
phase, in which the flux lines are localized on the line
defects.

Although our model is presumably not in the
same universality class (because of the long range interactions among the
vortices) as that
considered by Nelson and Vinokur, we briefly interpret our results in the
context of this theory.  The ``Bose glass'' regime, in which
the line defect density f$_d > f$, corresponds to our Model A and one of
the cases considered in our Model B.  In this regime, for
T $< T_c(B)$, Vinokur and Nelson
predict a nonlinear IV relation of the form
V/J $\propto$ exp$[-(E_k/k_BT)(J_0/J)^{1/3}]$ where J is the current
density and $E_k$ and $J_0$ are constants depending on the strength of the
pinning potential and other parameters.  Our IV characteristics for Model
A (with $\vec{J} \perp \vec{B}$)
are not inconsistent with this behavior, though our samples
are extremely small in the dynamical calculations. (Although the number of
flux {\it lines} is very small in these calculations, the number of
flux {\it loops}, formed by localized bowing out of the flux lines in
response to applied currents, is considerably larger.  Since dissipation
occurs largely by motion of these loops, in the Bose glass model, our
dynamical samples may not be as inadequately small as appears on first
sight.) For $\vec{J} \perp \vec{B}$, a simple activation
form [V/J $\propto \exp(-E_kJ_0/(Jk_BT))$] seems ruled out by our
results.  For $\vec{J} \| \vec{B}$, our data seems consistent with an
activated behavior of this form; however,
we are not aware of an analytical theory
describing Bose glass transport in this regime. Obviously,
more detailed numerical simulations, involving
much larger numbers of line defects and more disorder realizations,
are needed before any definite conclusions can be drawn
about the quantitative applicability of the Bose glass
picture to this dynamical model.

Our results for Model A disorder seem consistent with both static
and dynamic scaling hypotheses as applied to this anisotropic phase
transition, with critical indices
$z \approx 1.4 \pm 0.2, z^{\prime} \approx 1.8, \nu_{\perp} \approx 1.3\pm
0.3$.  z is, as expected, smaller than that found in analogous calculations
with
short-range interactions$^{26}$ between
vortex lines, for which z = 2, but is, perhaps coincidentally, in
the range of findings for other lattice models with line
disorder$^{29}$.  Our results (although based on very small
dynamical samples) seem to suggest there may be only a
single dynamical exponent z$^{\prime}$
for transport both perpendicular and parallel to the line defects, in
contrast to the short-range model.  This conclusion may, however, be a
function of the particular dynamics assumed.$^{30}$
The value of $\nu_{\perp}$ is
comparable to values of $\nu$ which have been reported in the experimental
literature$^{31}$ in cases where point disorder is expected to dominate; to our
knowledge, no direct measurement of $\nu_{\perp}$ has been carried out for
samples with known line disorder.

To summarize, we have studied flux pinning by defects in three-dimensional
Josephson-junction networks with various types of point and line defects.
We find that line defects considerably enhance both the critical current
and the upper critical field, relative to the same concentration of point
defects.  We also find some indications that the normal-to-superconducting
transition in the case of random line disorder is a continuous phase
transition marked by both static and dynamical critical phenomena.
These conclusions are, however,
subject to large numerical uncertainties arising from the
relatively small samples studied.  The resulting
critical exponents may be applicable to high-T$_c$ superconductors in which
the intrinsic anisotropy is not too great
(such as YBa$_2$Cu$_3$O$_{7-\delta}$) and if screening fields can be
neglected (extreme Type II limit).   Larger scale calculations, including
screening, and making use of more realistic pinning models, will be necessary
for quantitative comparisons to experiment.  These are planned for future
work.

\vspace{0.2in}
\noindent
{\large \bf Acknowledgments}

\vspace{0.05in}
\noindent
We should like to thank M. P. A. Fisher, D. A. Huse, M. Makivic,
D. R. Nelson, Mats Wallin, and A. P. Young for many useful conversations.
This work has been
supported by the Midwest Superconductivity Consortium (MISCON) through
Department of Energy grant DE-FG-45427, and in part by
NSF grant DMR 90-20994.  Calculations were carried out on the
CRAY Y-MP 8/864 of the Ohio Supercomputer Center.

\vspace{0.5cm}

\noindent
{\bf Appendix A. Static Scaling.}

We describe our numerical results for line defects within the framework
of a scaling analysis suitable for both our static and dynamic
results, based largely on previous discussions of Nelson
and Vinokur.$^{21}$
In this Appendix, we describe the static scaling hypotheses.

Consider a HTSC with line defects and a magnetic field both
oriented in the z direction.  Suppose that there is a phase transition at
some temperature T$_c$(B), where B is the magnitude of the applied magnetic
field.  Assume also that this phase transition is characterized by
{\it two} diverging correlation lengths,
$\xi_{\perp}$ and $\xi_{\|}$, corresponding to correlations in the xy plane
and z direction, respectively.  To allow for the possibility that these
diverge with different critical exponents, we write
\begin{eqnarray}
\xi_{\perp} \propto t^{-\nu_{\perp}} \\
\xi_{\|} \propto t^{-z\nu_{\perp}} \\
t = |T-T_c(B)|/T_c(B).
\end{eqnarray}
An isotropic phase transition is a special case of this behavior with z = 1.

By analogy with the usual isotropic
hyperscaling expression for the singular part $f_s$
of the free energy density near T$_c$, we assume that f$_s$ behaves as
\begin{equation}
\beta f_s \propto \frac{1}{\xi_{\perp}^2\xi_{\|}}
\end{equation}
(where $\beta = 1/k_BT$).
Hence, the specific heat has a singularity of the form
\begin{equation}
C_V \propto \frac{\partial^2 f}{\partial t^2}
\propto t^{\nu_{\perp}(2+z)-2}.
\end{equation}

To estimate the behavior of $\gamma_{\perp}$ and $\gamma_{\|}$, the
principal components of the helicity modulus tensor, we extend an
argument of Cha {\it et al}$^{23}$ to anisotropic phase transitions.
First imagine that the array is subjected to a phase gradient
$\vec{\nabla}_z\theta$ in the z direction.  The change in free energy
per unit volume is
\begin{equation}
\beta f_s \propto \frac{1}{2}\gamma_{\|}|\vec{\nabla_z}\theta|^2
\propto \frac{\gamma_{\|}}{\xi_{\|}^2},
\end{equation}
where we have replaced $\vec{\nabla}_z$ by the inverse of the
characteristic length $\xi_{\|}$.  With the use of eqs.(18) and (20), this
gives
\begin{equation}
\gamma_{\|} \propto \frac{\xi_{\|}}{\xi_{\perp}^2} \propto
t^{(2-z)\nu_{\perp}},
\end{equation}
where the last proportionality describes the expected critical behavior
near T$_c$.
A similar argument applied to $\gamma_{\perp}$ gives
\begin{equation}
\gamma_{\perp} \propto \frac{1}{\xi_{\|}} \propto t^{z\nu_{\perp}}.
\end{equation}

In a Monte Carlo calculation, it is necessary to calculate these quantities
in finite-size sample, usually a parallelopiped of volume
L$_{\perp}^2L_{\|}$.  The natural scaling form for the helicity moduli in
such samples is
\begin{eqnarray}
\gamma_{\perp} = \frac{1}{\xi_{\|}}F(\xi_{\|}/L_{\|}, \xi_{\perp}/L_{\perp})
 \\
\gamma_{\|} = \frac{\xi_{\|}}{\xi_{\perp}^2}G(\xi_{\|}/L_{\|},\xi_{\perp}/
L_{\perp}),
\end{eqnarray}
where F(u, v) and G(u, v) are universal functions.  Expressions (25) and
(26) can be written in more convenient forms by
making the change of variables $F(u, v) = uH(uv^{-z}, u)$,
$G(u, v) =(v^2/u)K(uv^{-z}, u)$ to yield
\begin{eqnarray}
\gamma_{\perp} = \frac{1}{L_{\|}}H(L_{\perp}^z/L_{\|}, L_{\|}/\xi_{\|})\\
\gamma_{\|} = \frac{L_{\|}}{L_{\perp}^2}K(L_{\perp}^z/L_{\|}, L_{\|}/
\xi_{\|}).
\end{eqnarray}
At T = T$_c$, the second argument of both H and K vanishes.  Hence, for a
finite sample whose dimensions are chosen such that
L$_{\|} \propto L_{\perp}^z$, it follows that
\begin{eqnarray}
L_{\|}\gamma_{\perp} = const.; \\
L_{\perp}^2\gamma_{\|}/L_{\|} = const.
\end{eqnarray}
These relations can, in principle,
be used to determine the transition temperature with
high accuracy by examining the behavior of the components of $\gamma$ in
a series of boxes of different volumes, such that the ratio
L$_{\|}/L_{\perp}^z$ is held constant.  The method is to plot
L$_{\|}\gamma_{\perp}$ and L$_{\perp}^2\gamma_{\|}/L_{\|}$ for different
volumes; all should cross at T = T$_c$.  Unfortunately, this method works
only {\it provided z is known}.  Since z is apparently in the
range 1.2 - 1.5 for the present model, but difficult to determine with
greater accuracy, we have not attempted this kind of
anisotropic finite-size scaling in the present paper.

\vspace{0.1in}

\noindent
{\bf Appendix B. Dynamic Scaling.}

For dynamical quantities, we may again follow and somewhat extend
the arguments of Nelson and Vinokur.$^{21}$  We consider first the
electric field
E$_{\perp}$ and current density J$_{\perp}$ in the transverse direction.
In this case, we postulate a scaling relation of the form
\begin{equation}
E_{\perp} = \xi_{\perp}^{-a}
E_{\pm, \perp}(\hbar J_{\perp}\xi_{\perp}\xi_{\|}/(2ek_BT))
\end{equation}
where E$_{\pm, \perp}$ are scaling functions which apply respectively above
and below T$_c$.  To determine $a$, suppose the HTSC is in the Ohmic regime
at T $> T_c$.  In this regime, E$_{+,\perp}(x) \propto x$, whence
$\sigma_{\perp} \equiv J_{\perp}/E_{\perp} \propto \xi_{\perp}^{a-1-z}$.
However, we also expect that
$\sigma_{\perp}$ should scale like $\gamma_{\perp}/\omega$ (that is, both
of these quantities should have the same power-law dependence on
$\xi_{\perp}$, where $\omega$ is a characteristic frequency).
In turn, we expect that $\omega$ should vanish near this continuous phase
transition, with a characteristic temperature-dependence given by
$\xi_{\perp}^{-z'}$ where $z'$ is a new dynamical critical exponent.
Combining all these relations, and using eq. (24), we obtain
\begin{equation}
a = 1 + z'.
\end{equation}

Similarly, for transport parallel to the z axis, we expect the scaling
relation
\begin{equation}
E_{\|}= \xi_{\perp}^{-b}E_{\pm, \|}(\hbar J_{\|}\xi_{\perp}^2/(2ek_BT)),
\end{equation}
and making arguments analogous to the perpendicular case, we find
\begin{equation}
b = z + z'.
\end{equation}

Precisely at T = T$_c$ we expect both E$_{\perp}$ and E$_{\|}$ to vary as
power laws in J$_{\perp}$ and J$_{\|}$ respectively.  This behavior
implies that at T$_c$ the scaling functions E$_{\pm,\perp}(x)$ and
E$_{\pm,\|}(x)$ should take the forms
\begin{eqnarray}
E_{\pm, \perp}(x) \propto x^c, \\
E_{\pm, \|}(x) \propto x^d.
\end{eqnarray}
The exponents $c$ and $d$ can be determined by
observing that $\xi_{\perp}$ and $\xi_{\|}$ are infinite
at T$_c$.  In order for eqs. (31) and
(33) still to be satisfied at T$_c$, the left and right hand sides must
should involve equal powers of $t$.  This leads to the results
\begin{eqnarray}
c = (1+z')/(1+z) \\
d = (z+z')/2.
\end{eqnarray}
Thus, calculating or measuring
the current-voltage characteristics precisely at T=T$_c$
should yield power laws whose slopes determine
the exponents $z$ and $z'$.

The above arguments assume that there is a {\it single} dynamical
critical exponent z$^{\prime}$.  If instead, there are two such exponents
z$_{\perp}^{\prime}$ and z$_{\|}^{\prime}$ describing the divergent
relaxation times in the perpendicular and parallel directions, then eqs.
(37) and (38) are replaced by the relations
c = (1+z$_{\perp}^{\prime}$)/(1+z); d = (z+z$_{\|}^{\prime}$)/2.

\vspace{0.1in}

\noindent
{\bf Appendix C.  Dynamic Scaling in Finite Size Systems.}

Our finite-size IV characteristics often show an Ohmic tail at low currents,
even at temperatures well below the putative superconducting transition.
In this Appendix, we give an argument suggesting that this tail is a
finite-size effect which would disappear in sufficiently large systems.

We present the argument for the case J${\|}$B, where the numerical results
most clearly show the finite-size tail.  However, a similar argument should
also hold for the perpendicular case.  In the parallel case, for T$<T_c$ in
an infinite system, as shown in
Appendix B, the electric field and current density are related by
\begin{equation}
E_{\|} =  \xi_{\perp}^{-(z+z')}E_{-, \|}
(\hbar J_{\|}\xi_{\perp}^2/(2ek_BT)),
\end{equation}
where E$_{-}$ is some universal function.
Now write E$_{-,\|}(x) = xF_{-,\|}$(x).  It follows that the resistivity
$\rho_{\|} \equiv E_{\|}/J_{\|}$ can be written as
\begin{equation}
\rho_{\|} \propto \xi_{\|}^{2-z-z'}F_{-, \|}(\hbar
J_{\|}\xi_{\perp}^2/(2ek_BT)).
\end{equation}

For a cubic system of edge L, this relation must involve another variable,
the ratio $\xi_{\perp}/L$:
\begin{equation}
\rho_{\|} \propto \xi_{\perp}^{2-z-z'}
F_{-,\|}(\hbar J_{\|}\xi_{\perp}^2/(2ek_BT), \xi_{\perp}/L).
\end{equation}

The numerical data presented in Fig. 7(b) suggest that
 $F_{-, \|}(x, 0)$ falls rapidly to zero with decreasing x.  Indeed, the
data seems to fit roughly to the relation
\begin{equation}
F_{-,\|}(x, 0) = F_o\exp(-A/x^{\mu_{\|}})
\end{equation}
where $F_o$ and A are constants and $\mu_{\|} \approx 1)$.  We have no
theory for the finite-size version of this function, but a plausible guess
suggests itself.  The dimensionless
argument $x = \hbar J_{\|}\xi_{\perp}^2/(2ek_BT)$ may be expressed as
$x = (\xi_{\perp}/\xi_J)^2$ where $\xi_J = (2ek_BT/(\hbar J_{\|}))^{1/2}$
is a characteristic length defined by the current density J$_{\|}$ (note
that length is measured in units of the intergranular separation).
When
this length becomes larger than the system size, the current length should
be replaced by L.  In order to include this behavior in the scaling form,
we may postulate
\begin{eqnarray}
F_{-, \|}(\hbar J_{\|}\xi_{\perp}^2/(2ek_BT), \xi_{\perp}/L)\nonumber \\
         =F_o\exp(-A/[\hbar J_{\|}\xi_{\perp}^2/(2ek_BT)+\xi_{\perp}^2/L^2]
^{\mu_{\|}}).
\end{eqnarray}
This function has several desirable properties.  First, for large L, it
reduces to the infinite-size form of eq. (42).  Secondly, for very small
J$_{\|}$, $\rho_{\|}$ becomes independent of $J_{\|}$ (i. e., becomes
Ohmic) and given by
\begin{equation}
\rho_{\|} \propto \xi_{\|}^{2-z-z'}\exp(-AL^{2\mu_{\|}}/\xi_{\|}^{2\mu_{\|}}).
\end{equation}

Eq. (44) is the desired low-current Ohmic tail seen in our calculations.
As expected, it goes away at large enough sizes, or low enough temperatures
($\xi_{\|}$ becomes smaller and smaller as the temperature is decreased
{\it below} T$_c$).  Since z + z$^{\prime} \approx 4 - 5$, the
prefactor in eq. (41) grows with decreasing temperature.  However, its
growth should be more than offset by the decreasing exponential, so that
$\rho_{\|}$ should decrease with decreasing temperature for fixed large L.
Thus, the argument presented in this Appendix
gives a plausible explanation of the finite-size numerical results
discussed in the text.

\newpage

\begin{figure}
\caption{
Specific heat per grain, C$_V$; and parallel and perpendicular components
of the helicity moduli, $\gamma_{\|}$ and $\gamma_{\perp}$,
for an ordered L $\times$ L $\times$ L lattice
of grains, with isotropic coupling ($E_{J,\perp} = E_{J,\|} \equiv E_J$),
magnetic field f = 1/4, and L = 8, 12, and 16, with periodic
boundary conditions, plotted as a function of temperature T.}
\end{figure}

\begin{figure}
\caption{
L$\gamma_{\|}$ and L$\gamma_{\perp}$
versus temperature, for the array of Fig. 1.}
\end{figure}

\begin{figure}
\caption{
Same as Fig. 1, but for an array with Model A line disorder in the bond
strengths.  The calculations involve averages over 35, 21, and 19
realizations of the disorder for L = 8, 12, and 16 respectively. }
\end{figure}

\begin{figure}
\caption{
IV characteristics for the model of Fig. 1, L $\times$ L $\times$ L$_z$ array,
current density $\vec{J} \perp \vec{B}$, at several different temperatures,
averaged over three different (random) choices of initial conditions, and L
= 6, L$_z$ = 9.
In this and subsequent Figures, the lines are simply interpolations between
calculated points.}
\end{figure}

\begin{figure}
\caption{
IV characteristics for the model of Fig. 3, L $\times$ L $\times$ L array,
with L = 6:
(a) $\vec{J} \perp \vec{B}$, and (b)
$\vec{J} \| \vec{B}$, at several different temperatures, as indicated.}
\end{figure}

\begin{figure}
\caption{
Resistivity $\rho \equiv <V>/(LRI)$ for an L $\times$ L $\times$ L array
with L = 6, at a current level I = 0.05I$_c$ per grain, for f = 1/4 and
no defects, $\vec{J} \perp \vec{B}$ (triangles); line
defects, $\vec{J}\perp\vec{B}$ (squares); and line defects,
$\vec{J} \| \vec{B}$ (circles).  Cases (b) and (c) each
involve averages over ten realizations of the disorder.}
\end{figure}

\begin{figure}
\caption{
Scaling plots of the IV characteristics from Figs. 5 for (a)
$\vec{J} \perp \vec{B}$ and (b) $\vec{J} \| \vec{B}$.  In both
cases, the IV characteristics both above and below T$_c(B)$ collapse
reasonably well onto universal scaling functions over the limited current
ranges considered.  For T $<$ T$_c$, there are low-current Ohmic tails in
both cases (especially for $\vec{J} \| \vec{B}$), which are probably due to
finite size effects.  The fitting parameters for curves (a) are z =1.5,
z$^{\prime}=2.0$, $\nu_{\perp}=1.5$, k$_BT_c=1.7E_J$; for curves (b), they
are z=1.5, z$^{\prime}=1.3$, $\nu_{\perp}=1.2$, and k$_BT_c$=1.7E$_J$. }
\end{figure}

\begin{figure}
\caption{
IV characteristics for Model B disorder and $\vec{J}\perp\vec{B}$,
at temperature T = 0,
an L $\times$ L $\times$ L$_z$ array, with L = 8, L$_z=5$,
magnetic flux f = 1/8 with
$\vec{B}$ parallel to the z (thin)direction:
no defects (dotted curve); 40 randomly distributed
point defects (full curve, average of 7 realizations);
8 randomly distributed line defects parallel to the z
direction (dashed curve, average of 10 realizations); and 8 periodically
distributed line defects in the z direction (dot-dashed curve).
I is the applied current per grain; $<V>$ is the time-averaged voltage across
the sample, averaged over the directions perpendicular to the current; R is
the shunt resistance, and I$_c$ is the critical current of each junction.
For reference, the critical current for the ordered lattice at f = 0 is
I/I$_c$=1.0.  }
\end{figure}

\begin{figure}
\caption{
Resistivity $\rho(T) \equiv$ V/I, at an applied current I = 0.1I$_c$ per
junction, plotted versus temperature T, at f = 1/8.  Symbols as in Fig.
10.  Inset: ground state vortex line configuration for f=1/8 lattice.
Filled squares denote loci of vortex lines (plaquettes of positive vorticity,
i.e. current circulating counterclockwise);
empty squares are plaquettes of negative vorticity.  For reference,
T$_c(f=0) \approx 2.21E_J/k_B$.}
\end{figure}

\begin{figure}
\caption{
Same as Fig. 8, but for a flux f = 1/8 and several densities f$_d$
of columnar defects oriented parallel to the z axis.  Each curve represents
an average over ten realizations of the disorder.
In this case, the lattice is $8 \times$ 8 $\times$ 8.}
\end{figure}

\begin{figure}
\caption{
Same as Fig. 9, but for a flux f = 1/8 and several densities f$_d$ of
columnar defects oriented parallel to the z axis.  Each curve represents an
average over ten realizations of the disorder.}
\end{figure}

\end{document}